\documentclass[12pt]{article}
\usepackage{amssymb,amsmath,epsfig}

\begin{document}

\title{\bf Re-Scaling of Energy in the Stringy Charged Black Hole Solutions using
Approximate Symmetries}

\author{M. Sharif \thanks{msharif@math.pu.edu.pk} and Saira Waheed
\thanks{sairawaheed\_50@yahoo.com}\\
Department of Mathematics, University of the Punjab,\\
Quaid-e-Azam Campus, Lahore-54590, Pakistan.}

\date{}

\maketitle
\begin{abstract}
This paper is devoted to study the energy problem in general
relativity using approximate Lie symmetry methods for differential
equations. We evaluate second-order approximate symmetries of the
geodesic equations for the stringy charged black hole solutions. It
is concluded that energy must be re-scaled by some factor in the
second-order approximation.
\end{abstract}

\section{Introduction}

One of the most important issues of General Relativity (GR) is the
localization of energy-momentum. It is a conserved quantity in
classical mechanics. In GR, energy conservation is guaranteed only
for those spacetimes which are time translational invariant. In that
case, energy is defined as the dot product of timelike Killing
vector $\textbf{k}$ and momentum $4$-vector $\textbf{p}$, i.e.,
$E=\textbf{k}.\textbf{p}$. However, there are numerous spacetimes
which are non-stationary (admitting no timelike isometry and
consequently energy is not conserved there) \cite{1}. For the case
of gravitational waves, this problem is of special interest as
$T_{\mu\nu}=0$ and hence energy is not given by the energy-momentum
tensor. The existence of gravitational waves was demonstrated by
Weber and Wheler \cite{2} and Ehlers and Kundt \cite{3}. They
proposed an approximate formula to check the momentum imparted by
gravitational waves to test particles placed in their path. Using
the pseudo-Newtonian formalism, an exact formula was presented by
Qadir and Sharif \cite{4}. They showed that the results obtained for
plane and cylindrical gravitational waves coincide with those found
by the above people.

In order to have a trim expression for energy and momentum in GR,
Einstein \cite{5} himself proposed a prescription which is a
combination of the energy-momentum tensor and the pseudo-tensor.
Many others including Landau-Lifshitz, Papapetrou, Weinberg etc.
\cite{6}-\cite{8} also developed energy-momentum complexes. All
these prescriptions are coordinate dependent providing meaningful
results only in Cartesian coordinates. In order to overcome this
difficulty, M$\o$ller, Bondi, Ashtekar-Hansen \cite{9}-\cite{11} and
others formulated coordinate independent prescriptions. It has been
shown that for a given spacetime, various prescriptions not
necessarily provide the same result. However, none of these attempts
proved to be unambiguous and consequently could not yield a correct
definition. Therefore it would be interesting to develop some
procedure which should be more accurate than the previous ones.

The spacetime symmetries are characterized by its isometries which
form a Lie algebra \cite{17}. The symmetries of differential
equations (DEs) also form Lie algebra \cite{18}. Since the geodesic
equations are the DEs therefore it has been suggested that there
would be a nexus between spacetime isometries and the symmetries of
the geodesic equations \cite{19,20}. The symmetries of the manifold
which are obtained through the geodesic equations yield some
additional symmetries which provide no conservation law. However,
other symmetries yield quantities conserved under geodesic motion as
well as first integrals of the geodesic equations \cite{21,22}. A
manifold with no exact symmetry can possess approximate symmetries
and more interesting information can be obtained from a slightly
broken (approximate) symmetry than the exact symmetry. There are
different methods available to find approximate symmetries of DEs.
Two ``approximate symmetry" theories have been constructed by using
the combination of Lie group theory and perturbations. The first
theory was proposed by Baikov et al. \cite{23} and the second theory
was given by Fushchich and Shtelen \cite{24}.

Kara et al. \cite{25} used the first approximate symmetry method
to discuss conservation laws of energy and momentum for the
Schwarzschild spacetime. Later, this procedure was applied to the
Reissner-Nordstr$\ddot{o}$m (RN) solution \cite{26} and an energy
re-scaling factor was obtained in the second-order approximation.
Hussain et al. \cite{27,28} found energy re-scaling factors for
the Kerr-Newmann and Kerr-Newmann AdS spacetimes by using the same
procedure.

In going from the Minkowski spacetime (flat) to non-flat stringy
charged solutions, the recovered approximate symmetries and
consequently the conservation laws (recovered in the first-order
approximation) are lost. It is expected that in the limit of small
charge we should retrieve all the lost symmetries. For this purpose,
we have to find out the second order approximate symmetries. In this
paper, we use the first procedure of approximate symmetry to
evaluate approximate symmetries and to obtain energy re-scaling
factor for the stringy charged black holes.

The paper is organized as follows. In the next section, we discuss
exact and approximate symmetry methods for the solution of DEs.
Section \textbf{3} contains exact symmetries of the Minkowski
spacetime and the first-order approximate symmetries of the
Schwarzschild spacetime. In section \textbf{4}, we study
approximate symmetries of the stringy charged black hole
solutions, i.e., electrically and magnetically charged solutions.
Finally, we summarize and discuss the results in the last section.

\section{Mathematical Formulation}

Here we review approximate symmetry analysis for the solution of
DEs. Symmetry is a point transformation, it may be independent of
the choice of variables, which maps the solutions into the
solutions. In other words, one can say that symmetries are the
transformations under which the form of DEs does not change. A
point transformation is a transformation which maps one point
$(x,y)$ into another point $(x^{*},y^{*})$. Symmetries are very
useful in finding the solution of DEs or reducing them to the more
simpler form for integration. For example, with the help of
symmetries, one can convert non-linear DEs which arise in many
practical problems into linear DEs. Also, the importance of
symmetries lies in the most celebrated Noether's theorem which is
stated as ``Any differentiable symmetry of the action of a
physical system has a corresponding conservation law"
\cite{29,30}. If, for a given system of DEs, there is a
variational principle, then a continuous symmetry invariant under
the action of the functional provides a conservation law
\cite{31}-\cite{33}.

The symmetry generator of a $n^{th}$-order ordinary differential
equation (ODE) (involving $s$ as independent variable and
$\textbf{x}$ as dependent variable and
$\textbf{x}',~\textbf{x}'',...,~\textbf{x}^{(n)}$ represents its
first, second and so on $n^{th}$-order derivative with respect to
$s$) \cite{18,34}
\begin{equation}\label{1}
\textbf{E}(s;\textbf{x}(s),\textbf{x}'(s),\textbf{x}''(s),...,\textbf{x}^{(n)}(s))=0
\end{equation}
under the point transformation: $(s,\textbf{x})\longrightarrow
(\xi(s,\textbf{x}),\eta(s,\textbf{x}))$ can be found if on the
solution of the ODE, $\textbf{E}=0$, the following condition is
satisfied
\begin{equation*}
\textbf{X}^{[n]}(\textbf{E})\mid_{\textbf{E}=0}=0.
\end{equation*}
Here, $\textbf{X}^{[n]}$ is the $n^{th}$-order prolongation or
extension of the infinitesimal symmetry generator
\begin{equation}\label{4}
\textbf{X}=\xi(s,\textbf{x})\frac{\partial}{\partial
s}+\eta(s,\textbf{x})\frac{\partial}{\partial \textbf{x}}
\end{equation}
and is given by
\begin{eqnarray}\label{2}
\textbf{X}^{[n]}&=&\xi(s,\textbf{x})\frac{\partial}{\partial
s}+\eta(s,\textbf{x})\frac{\partial}{\partial
\textbf{x}}+\eta_{,s}(s,\textbf{x},\textbf{x}')\frac{\partial}{\partial \textbf{x}'}+...\nonumber\\
&+&\eta_{,(n)}(s,\textbf{x},\textbf{x}',...,\textbf{x}^{(n)})\frac{\partial}{\partial
\textbf{x}^{(n)}}.
\end{eqnarray}
The values of the prolongation coefficients are given by
\begin{eqnarray}\label{10}
\eta_{,s}=\frac{d\eta}{ds}-\textbf{x}'\frac{d\xi}{ds},\quad
\eta_{,(n)}=\frac{d\eta_{,(n-1)}}{ds}-\textbf{x}^{(n)}\frac{d\xi}{ds};\quad
n\geq2.
\end{eqnarray}

In the same manner, the system of $p$ ODEs of order $n$
\begin{equation}\label{5}
\textbf{E}_{\alpha}(s;\textbf{x}(s),\textbf{x}'(s),\textbf{x}''(s),...,\textbf{x}^{(n)}(s))=0,\quad
(\alpha=1,2,3..,p)
\end{equation}
admits a symmetry generator
\begin{equation}\label{3}
\textbf{X}=\xi(s,\textbf{x})\frac{\partial}{\partial
s}+\eta^{\alpha}(s,\textbf{x})\frac{\partial}{\partial
\textbf{x}^{\alpha}}
\end{equation}
if and only if for each ODE, the following symmetry condition
\begin{equation*}
\textbf{X}^{[n]}(\textbf{E}_{\alpha})\mid_{\textbf{E}_{\alpha=0}}=0
\end{equation*}
is satisfied. In that case, $n^{th}$-order extension of the symmetry
generator (\ref{3}) is given by
\begin{eqnarray}\label{8}
\textbf{X}^{[n]}&=&\xi(s,\textbf{x})\frac{\partial}{\partial
s}+\eta^{\alpha}(s,\textbf{x})\frac{\partial}{\partial
\textbf{x}}+\eta^{\alpha}_{,s}(s,\textbf{x},\textbf{x}')\frac{\partial}{\partial
\textbf{x}^{\alpha'}
}+...\nonumber\\
&+&\eta^{\alpha}_{,(n)}(s,\textbf{x},\textbf{x}',...,\textbf{x}^{(n)})\frac{\partial}{\partial
\textbf{x}^{\alpha(n)}}
\end{eqnarray}
and the corresponding prolongation coefficients are
\begin{eqnarray}\label{9}
\eta^{\alpha}_{,s}=\frac{d\eta^{\alpha}}{ds}-\textbf{x}^{\alpha'}\frac{d\xi}{ds},\quad
\eta^{\alpha}_{,(n)}=\frac{d\eta^{\alpha}_{,(n-1)}}{ds}-\textbf{x}^{\alpha(n)}\frac{d\xi}{ds};\quad
n\geq2.
\end{eqnarray}

If $p=1$, then the system reduces to a single equation. The
$k^{th}$-order approximate symmetry of a perturbed system of ODEs
\begin{equation}\label{11}
\textbf{E}=\textbf{E}_{0}+{\epsilon}\textbf{E}_{1}+\epsilon^{2}\textbf{E}_{2}+...+\epsilon^{k}\textbf{E}_{k}
+O(\epsilon^{k+1})
\end{equation}
is given by the generator
\begin{equation}\label{12}
\textbf{X}=\textbf{X}_{0}+{\epsilon}\textbf{X}_{1}+\epsilon^{2}\textbf{X}_{2}+...+\epsilon^{k}\textbf{X}_{k}
\end{equation}
if the following symmetry condition holds \cite{35}
\begin{eqnarray}\label{13}
\textbf{X}\textbf{E}&=&[(\textbf{X}=\textbf{X}_{0}+{\epsilon}\textbf{X}_{1}+\epsilon^{2}\textbf{X}_{2}+...+
\epsilon^{k}\textbf{X}_{k})
(\textbf{E}=\textbf{E}_{0}+{\epsilon}\textbf{E}_{1}\nonumber\\
&+&\epsilon^{2}\textbf{E}_{2}+...+\epsilon^{k}\textbf{E}_{k})]
\mid_{{\textbf{E}=\textbf{E}_{0}+{\epsilon}\textbf{E}_{1}+\epsilon^{2}\textbf{E}_{2}+...
+\epsilon^{k}\textbf{E}_{k}}} = O({\epsilon}^{k+1}).
\end{eqnarray}
Here $\epsilon\in R^{+},~\textbf{E}_{0}$ is the exact system of
equations, $\textbf{E}_{1},~\textbf{E}_{2}$ are the first and second
order perturbed parts of the perturbed DEs respectively and so on.
$\textbf{X}_{0}$ represents the exact part of the symmetry generator
and $\textbf{X}_{1},~\textbf{X}_{2}$ denote respectively the first
and second order approximate parts of the symmetry generator and so
on. For the $k^{th}$-order approximate symmetry generator, we put
terms involving $\epsilon^{(k+1)}$ and its higher powers equal to
zero (then the RHS of Eq.(\ref{13}) becomes zero). We know that
symmetries of an ODE always form a Lie algebra. However, the
approximate symmetries do not necessarily form a Lie algebra rather
do form the so-called "approximate Lie algebra" (up to a specified
degree of precision) \cite{36}. The perturbed equation always admits
an approximate symmetry $\epsilon \textbf{X}_{0}$, called a trivial
symmetry. If a symmetry generator
$\textbf{X}=\textbf{X}_{0}+\epsilon \textbf{X}_{1}$ exists with
$\textbf{X}_{0}\neq 0$ and $\textbf{X}_{1}\neq \textbf{X}_{0}$, then
it will be called non-trivial symmetry generator \cite{37}.

\section{Approximate Symmetries of the Schwarzschild Spacetime}

In this section, starting from the Minkowski spacetime which is
maximally symmetric with $10$ Killing vectors, we shall discuss
approximate symmetries of the Schwarzschild spacetime. The
isometries of the Minkowski spacetime form Poincare algebra,
$so(1,3)\oplus_{s}R^{4}$, where the semi-direct sum $\oplus_{s}$
indicates that the sub-algebras do not commute. Also, the algebra
$so(1,3)$ is isomorphic to $so(3)\oplus_{s} so(3)$. In this
algebra, $so(3)\oplus_{s} so(3)\oplus_{s}R^{4}$, one of $so(3)$
gives conservation of angular momentum, another $so(3)$ provides
conservation of spin angular momentum, $R^{4}$ gives conservation
of energy and linear momentum. The symmetry generators of the
Minkowski spacetime are given \cite{38} in Appendix \textbf{A}.
The algebra calculated from the geodesic equations for the
Minkowski spacetime contains some symmetries which do not
correspond to any conservation law. These symmetries arise due to
re-mixing of the geodetic parameter with the Noether symmetries.

The Schwarzschild spacetime is given by
\begin{equation}\label{23}
ds^2=e^{\nu{(r)}}dt^2-e^{-\nu(r)}dr^2-r^2(d\theta^2+\sin^2\theta
d\phi^2),\quad e^{\nu{(r)}}=1-\frac{2M}{r}.
\end{equation}
This has four isometries
$\textbf{X}_{0},~\textbf{X}_{1},~\textbf{X}_{2},~\textbf{X}_{3}$
which provide conservation laws of energy and angular momentum.
However, the conservation laws of linear and spin angular momentum
are lost due to the presence of gravitational mass. The symmetry
algebra of the Schwarzschild spacetime (calculated through the
geodesic equations) consists of the four isometries and the dilation
algebra $d_{2}$ (algebra corresponding to $\frac{\partial}{\partial
s}$ and $s\frac{\partial}{\partial s}$), i.e., $so(3)\oplus R \oplus
d_{2}$ ($\oplus$ denotes the direct sum). In the limit of small mass
of point gravitating source, $\epsilon=2M$ \cite{25}, and using
approximate symmetry analysis, all lost conservation laws are
recovered as first-order trivial approximate symmetries yielding
approximate conservation laws of these quantities. Also, for the
more restricted equation of motion, i.e., the orbital equation,
along with the symmetries
\begin{eqnarray}\label{30}
\textbf{Y}_{0}&=&u\frac{\partial}{\partial u },\quad
\textbf{Y}_{1}=\cos\phi\frac{\partial}{\partial u},\quad
\textbf{Y}_{2}=\sin\phi\frac{\partial}{\partial u},\\\label{31}
\textbf{Y}_{3}&=&\frac{\partial}{\partial \phi },\quad
\textbf{Y}_{4}=\cos2\phi\frac{\partial}{\partial \phi}-u
\sin{2\phi}\frac{\partial}{\partial u},\\\label{32}
\textbf{Y}_{5}&=&\sin2\phi\frac{\partial}{\partial \phi}+u
\cos{2\phi}\frac{\partial}{{\partial}u},
\end{eqnarray}
\begin{eqnarray}\label{33}
\textbf{Y}_{6}&=&u\cos\phi\frac{\partial}{\partial \phi}-u^{2}
\sin\phi\frac{\partial}{\partial u},\\\label{34}
\textbf{Y}_{7}&=&u\sin\phi\frac{\partial}{\partial \phi}+u^{2}
\cos\phi\frac{\partial}{\partial u},
\end{eqnarray}
there exist two non-trivial stable approximate symmetries given by
\begin{eqnarray}\label{24}
\textbf{Y}_{a1}&=&\sin\phi \frac{\partial}{\partial
u}+{\epsilon}(2\sin\phi\frac{\partial}{\partial \phi}+u
\cos\phi\frac{\partial}{\partial u}) ,\\\label{25}
\textbf{Y}_{a2}&=&\cos\phi \frac{\partial}{\partial
u}-{\epsilon}(2\cos\phi\frac{\partial}{\partial \phi}-u
\sin\phi\frac{\partial}{\partial u}).
\end{eqnarray}

\section{Stringy Charged Black Holes}

In the framework of string theory, there  exist number of static
spherically symmetric black hole solutions. In this regard, the
first solution was given by Gibbons and Maeda \cite{39}. Later,
Garfinkle, Horowitz and Strominger developed electrically and
magnetically charged solutions (known as GHS solutions) \cite{40}.
Some other kinds of solutions were also found by different people.
Both string theory and GR have the same uncharged solutions. The
string analogues of the RN solution are obtained by solving low
energy string field equations. These string analogues of the RN
solution have causal structure of the Schwarzschild spacetime. In
$3+1$ dimensions, there exist two analogues of the RN spacetime:
one is associated with magnetic charge and another deals with the
electric charge. Both of these solutions are obtained by using the
solution generating techniques. We evaluate exact and approximate
symmetries of the geodesic and orbital equations of motion for
these GHS black hole solutions.

\subsection{Symmetries and Approximate Symmetries of the Stringy
Electrically Charged Black Hole}

The spacetime representing the electrically charged black hole
solution in string theory is given \cite{40} by
\begin{equation}\label{27}
ds^2=
\frac{(1-\frac{2m}{r})}{(1+\frac{2m\sinh^{2}\alpha}{r})^2}dt^2-\frac{1}
{(1-\frac{2m}{r})}dr^2-r^2(d\theta^2+\sin^2\theta d\phi^2).
\end{equation}
Here $m$ is the mass of the point gravitating source and $\alpha$ is
the electric charge of the black hole such that
$\tanh\alpha=\frac{Q^{2}}{ 2M^{2}}$. For $\alpha=0$, it reduces to
the Schwarzschild spacetime. For this spacetime, isometry algebra is
$so(3)\oplus R$ which corresponds to conservation laws of energy and
angular momentum while the conservation laws of linear and spin
angular momentum are lost. To recover these lost symmetries, we use
the approximate Lie symmetry method.

First, we concentrate to orbital equation of motion which is
obtained by restricting the motion of the particle to an
arbitrarily chosen equatorial plane. It is found \cite{25,26} that
there is a difference between the conservation laws obtained for
the full system of geodesic equations and the single orbital
equation of motion. To check whether this difference also holds in
the case of stringy electrically charged black hole solution, we
find the orbital equation of motion as follows
\begin{equation}\label{28}
\frac{d^{2}u}{d\phi^{2}}+u=\frac{m}{h^{2}}+3mu^{2}+\frac{2m\sinh^{2}\alpha}{h^{2}}
+\frac{4um^{2}\sinh^{4}\alpha}{h^{2}},
\end{equation}
where $h$ is the classical angular momentum per unit mass and
$u=\frac{1}{r}$. For the second order-approximate symmetries of
this equation, we define the perturbation parameter as
$\epsilon=2m$. Further, we assume
$m\sinh{\alpha}^{2}{\leq}{\epsilon}^{2}$ which implies that
$m\sinh^{2}\alpha=k\epsilon^{2}$, where $0<k\leq\frac{1}{4}$ (we
have taken these parameters because $\alpha=0$ leads to the
Schwarzschild spacetime and hence first-order perturbed spacetime
as first-order perturbed Schwarzschild spacetime and exact as
Minkowski spacetime). Introducing these perturbation parameters in
Eq.(\ref{27}) and retaining only the terms involving
$\epsilon^{2}$, the corresponding second-order perturbed spacetime
will be
\begin{equation*}
ds^2=(1-\frac{\epsilon}{r}-\frac{4k\epsilon^{2}}{r})dt^2
-(1+\frac{\epsilon}{r}+\frac{\epsilon^{2}}{r^{2}})dr^2
-r^2(d\theta^2+\sin^2\theta d\phi^2).
\end{equation*}
Similarly, the second-order perturbed orbital equation can be
written as
\begin{equation}\label{29}
\frac{d^{2}u}{d{\phi}^{2}}+u=\epsilon(\frac{1}{2h^{2}}+\frac{3}{2}
u^{2})+\epsilon^{2}(\frac{2k}{h^{2}}).
\end{equation}
For the exact symmetry generator, we put $\epsilon=0$ in
Eq.(\ref{29}) so that the exact symmetry generators are exactly the
same as given by Eqs.(\ref{30})-(\ref{34}).

In the first-order approximation when we retain only the terms
involving $\epsilon$ and neglect $O(\epsilon^{2})$, we get the
same two stable and non-trivial symmetry generators as given by
Eqs.(\ref{24}) and (\ref{25}) along with the symmetries given by
Eqs.(\ref{30})-(\ref{34}) (i.e. the approximate symmetries of the
Schwarzschild spacetime). For the second-order approximation, when
we apply the second-order prolonged operator $\textbf{X}^{[2]}$ to
the second-order perturbed orbital equation of motion,
\begin{equation*}
\textbf{E}:\quad
u''+u-\frac{\epsilon}{2h^{2}}(1+3u^{2}h^{2})-\epsilon^{2}(\frac{2k}{h^{2}})=0,
\end{equation*}
retaining only the terms involving $\epsilon^{2}$ and neglecting its
higher powers and substituting the values of prolongation
coefficients, we obtain a set of four DEs. The simultaneous solution
of these DEs shows that there exist some non-trivial symmetry
generators given as
\begin{eqnarray}\label{35}
\textbf{Y}_{a1}&=&u\frac{\partial}{\partial u }+\frac{2\epsilon
k}{h^{2}}\frac{\partial}{\partial u },\label{36}\\
\textbf{Y}_{a2}&=&\sin\phi\frac{\partial}{{\partial}u}+{\epsilon}(u\cos\phi\frac{\partial}{{\partial}
u}+2\sin\phi\frac{\partial}{{\partial}\phi}), \label{37}\\
\textbf{Y}_{a3}&=&\cos\phi\frac{\partial}{\partial
u}+{\epsilon}(u\sin\phi\frac{\partial}{\partial
u}-2\cos\phi\frac{\partial}{\partial\phi}), \label{38}\\
\textbf{Y}_{a4}&=&\cos2\phi \frac{\partial}{\partial \phi}-u
\sin2\phi\frac{\partial}{\partial u}-\frac{2k{\epsilon}}{h^{2}}
\sin2\phi\frac{\partial}{\partial u},
\\\label{39} \textbf{Y}_{a5}&=&\sin2\phi\frac{\partial}{\partial\phi}+u
\cos2\phi\frac{\partial}{{\partial}u}+\frac{2k\epsilon}{h^{2}}
\cos2\phi\frac{\partial}{{\partial}u},\label{40}\\
\textbf{Y}_{a6}&=&u\cos\phi\frac{\partial}{\partial \phi}-u^{2}
\sin\phi\frac{\partial}{\partial
u}-\epsilon(\frac{3ku}{h^{2}}\sin\phi\frac{\partial}{\partial u}),
\\\label{41}
\textbf{Y}_{a7}&=&u\sin\phi \frac{\partial}{\partial \phi}+u^{2}
\cos\phi\frac{\partial}{\partial
u}-\epsilon(\frac{3ku}{h^{2}}\cos\phi\frac{\partial}{\partial u}).
\end{eqnarray}
Since the re-scaling of energy of test particle was seen from the
approximate symmetries of the geodesic equations \cite{26}, thus
we apply this approximate symmetry analysis to the full system of
geodesic equations given by
\begin{eqnarray}\label{70}
&&\ddot{t}+2\frac{m[1+2\sinh^{2}\alpha-\frac{2m\sinh^{2}\alpha}{r}]}
{r^{2}(1-\frac{2m}{r})(1+\frac{2m\sinh^{2}\alpha}{r})}\dot{t}\dot{r}=0,\\
&&\ddot{r}+(1-\frac{2m}{r})[\frac{m}{r^{2}(1+\frac{2m\sinh^{2}\alpha}{r})^{2}}
+\frac{2m\sinh^{2}\alpha(1-\frac{2m}{r})}
{r^{2}(1+\frac{2m\sinh^{2}\alpha}{r})^{3}}]\dot{t}^{2}\nonumber\\\label{71}
&&-\frac{m}{r^{2}(1-\frac{2m}{r})}\dot{r}^{2}
-r(1-\frac{2m}{r})(\dot{\theta}^{2}+\sin^{2}\theta\dot{\phi}^{2})=0,
\end{eqnarray}
\begin{eqnarray}\label{72}
&&\ddot{{\theta}}+\frac{2}{r}\dot{r}\dot{\theta}- \sin\theta
\cos\theta \dot{\phi}^{2}=0,\\\label{73}
&&\ddot{{\phi}}+\frac{2}{r}\dot{r}\dot{\phi}+2\cot{\theta}
\dot{\theta}\dot{\phi}=0.
\end{eqnarray}

Now we introduce the perturbation parameters in the above
equations so that the corresponding perturbed geodesic equations
become
\begin{eqnarray}\label{74}
\textbf{E}_{1}: &&\ddot{t}+\frac{\epsilon}{r^{2}}\dot{t}\dot{r}
+\frac{{\epsilon}^{2}}{r^{3}}(1+4rk)\dot{t}\dot{r}=0, \\\nonumber
\textbf{E}_{2}:
&&\ddot{r}-r({\dot{\theta}}^{2}+{\sin{\theta}}^{2}{\dot{\phi}}^{2})
+{\epsilon}[\frac{1}{2r^{2}}({\dot{t}}^{2}-{\dot{r}}^{2})\\\label{75}
+&&({\dot{\theta}}^{2}+{\sin{\theta}}^{2}{\dot{\phi}}^{2})]
-{\frac{{\epsilon}^{2}}{2r^{3}}}[(1-4rk){\dot{t}}^{2}+{\dot{r}}^{2}]=0,\\\label{76}
\textbf{E}_{3}: &&\ddot{{\theta}}+\frac{2}{r}\dot{r}\dot{\theta}-
\sin\theta \cos\theta \dot{\phi}^{2}=0,\\\label{77} \textbf{E}_{4}:
&&\ddot{{\phi}}+\frac{2}{r}\dot{r}\dot{\phi}+2\cot{\theta}
\dot{\theta}\dot{\phi}=0.
\end{eqnarray}
Applying the second-order prolonged operator defined by
\begin{eqnarray}\nonumber
\textbf{X}^{[2]}&=& \xi\frac{\partial}{\partial
s}+\eta^{0}\frac{\partial}{\partial
t}+\eta^{1}\frac{\partial}{\partial
r}+\eta^{2}\frac{\partial}{\partial
\theta}+\eta^{3}\frac{\partial}{\partial
\phi}+\eta^{0}_{,s}\frac{\partial}{\partial
\dot{t}}+\eta^{1}_{,s}\frac{\partial}{\partial
\dot{r}}\\\label{80}&+&\eta^{2}_{,s}\frac{\partial}{\partial
\dot{\theta}}+\eta^{3}_{,s}\frac{\partial}{\partial \dot{\phi}}
+\eta^{0}_{,ss}\frac{\partial}{\partial
\ddot{t}}+\eta^{1}_{,ss}\frac{\partial}{\partial
\ddot{r}}+\eta^{2}_{,ss}\frac{\partial}{\partial
\ddot{\theta}}+\eta^{3}_{,ss}\frac{\partial}{\partial\ddot{\phi}}
\end{eqnarray}
to the perturbed geodesic equations, retaining only the terms
involving $\epsilon^{2}$ and substituting the values of the
prolongation coefficients, we obtain a system of $60$ DEs. In
construction of the system of determining equations for the
second-order approximation, we use Eqs.(\ref{14})-(\ref{16}) as
the four exact symmetry generators and remaining six as the
first-order approximate part of the symmetry generators. In this
set of determining equations, out of the four constants
corresponding to exact symmetry generators, two do not appear and
the remaining two cancel out. However, the six constants
corresponding to the first-order approximate part of the symmetry
generator are present. To make this system of DEs homogenous, we
have to eliminate these $6$ constants. The solution obtained by
back and forth substitution shows that all these constants vanish
and therefore the resulting system being similar to that of the
Minkowski spacetime, provides $12$ second-order approximate
symmetry generators. Among these generators, four are again exact
used earlier and they simply add into them while the remaining six
replace the lost first-order approximate symmetries. Thus the full
system has Poincare algebra $so(1,3)\oplus_{s}R^{4}$ and hence
there are no non-trivial second-order symmetries.

The exact symmetry generators consist of some symmetries associated
to dilation algebra, $\xi(s)=c_{0}s+c_{1}$. The terms involving
$\xi_{s}(s)=c_{0}$ cancel out in the set of equations for the
first-order approximate symmetries. However, these terms do not
cancel out automatically in the second-order approximation but
collect a scaling factor of $(1+4rk)$ for cancellation. We know that
energy conservation comes from the time translational invariance and
$\xi$ is the coefficient of $\frac{\partial}{\partial s}$ ($s$ is
the proper time) in the point transformation given by Eq.(\ref{3}).
Thus the coefficient of $c_{0}$ corresponds to energy re-scaling
factor given by
\begin{equation}\label{48}
(1+4rk)=1+\frac{r\sinh^2{\alpha}}{m}.
\end{equation}
Here we have used the value for $k$.

\subsection{Symmetries and Approximate Symmetries of
the Stringy Magnetically Charged Black Hole}

The spacetime representing the magnetically charged black hole
solution is \cite{40}
\begin{equation}\label{49}
ds^2=
\frac{(1-\frac{2M}{r})}{(1-\frac{Q^{2}}{Mr})}dt^2-\frac{1}{(1-\frac{2M}{r})
(1-\frac{Q^{2}}{Mr})}dr^2-r^2(d\theta^2+\sin^2\theta d\phi^2),
\end{equation}
where $M$ is the mass of the point gravitating source and $Q$ is
the magnetic charge of the black hole. For $Q=0$, the spacetime
reduces to the Schwarzschild solution. Here isometry algebra is
$so(3)\oplus R$ providing the conservation laws for energy and
angular momentum only. We discuss the exact and approximate
symmetries of this solution by applying the same procedure as
given in the previous section.

We discuss the symmetry structure of this spacetime for the
orbital equation of motion given by
\begin{equation}\label{50}
\frac{d^{2}u}{d{\phi}^{2}}+u=-\frac{Q^{2}}{Mh^{2}}(1-\frac{Q^{2}u}{M})+\frac{M}{h^{2}}-
\frac{2Q^{2}u}{h^{2}}+\frac{Q^{2}}{2Mh^{2}}+3Mu^{2}-4Q^{2}u^{3}+\frac{3}{2M}Q^{2}u^{2}.
\end{equation}
For the second order-approximate symmetries of this equation, we
define the perturbation parameters as
$\epsilon=2M,~Q^{2}/M=k\epsilon^{2}$, where we have assumed
$Q^{2}/M{\leq}\epsilon^{2}$ and therefore $k$ is given as
$0<k\leq\frac{1}{4}$ (again these perturbation parameters are
taken in order to reduce the perturbed spacetime for exact and
first-order approximation to Minkowski and first-order perturbed
Schwarzschild spacetimes respectively). Under these perturbations,
the second-order perturbed orbital equation is
\begin{equation}\label{52}
\frac{d^{2}u}{d{\phi}^{2}}+u=\epsilon(\frac{1}{2h^{2}}+\frac{3u^{2}}{2})
-\epsilon^{2}(\frac{k}{2h^{2}}-\frac{3ku^{2}}{2}).
\end{equation}
For the exact symmetry generator, we take $\epsilon=0$ in this
equation and the exact symmetries are given by
Eqs.(\ref{30})-(\ref{34}).

If we take only the terms involving $\epsilon$ and neglect the terms
of $\epsilon^{2}$ and its higher powers, i.e., the first-order
approximation, we get approximate symmetries of the Schwarzschild
spacetime including two non-trivial stable approximate symmetry
generators given by Eqs.(\ref{24}) and (\ref{25}). The second order
approximation yields no non-trivial symmetry generator. Only the
first-order symmetry generators are recovered. Hence there is no new
approximate conservation law but only the previous conservation laws
have been recovered.

The set of geodesic equations for this spacetime are
\begin{eqnarray}\label{84}
&&\ddot{t}+\frac{[\frac{2M}{r^{2}}-\frac{Q^{2}}{Mr^{2}}]}
{(1-\frac{2M}{r})(1-\frac{Q^{2}}{Mr})}\dot{t}\dot{r}=0,\\\nonumber
&&\ddot{r}+\frac{[\frac{2M}{r^{2}}-\frac{Q^{2}}{Mr^{2}}]
(1-\frac{2M}{r})}{2(1-\frac{Q^{2}}{Mr})}\dot{t}^{2}
+\frac{[\frac{2M}{r^{2}}-\frac{4Q^{2}}{r^{3}}+\frac{Q^{2}}{Mr^{2}}]}
{2(1-\frac{2M}{r})(1-\frac{Q^{2}}{Mr})}\dot{r}^{2}\\\label{85}
&&-r(1-\frac{2M}{r})(1-\frac{Q^{2}}{Mr})[\dot{\theta}^{2}
+\sin^{2}\theta\dot{\phi}^{2}]=0, \\\label{86}
&&\ddot{{\theta}}+\frac{2}{r}\dot{r}\dot{\theta}- \sin\theta
\cos\theta \dot{\phi}^{2}=0,\\\label{87}
&&\ddot{{\phi}}+\frac{2}{r}\dot{r}\dot{\phi}+2\cot{\theta}
\dot{\theta}\dot{\phi}=0.
\end{eqnarray}
Introduce the perturbation parameters defined earlier, the
corresponding second-order perturbed geodesic equations become
\begin{eqnarray}\label{60}
&&\ddot{t}+\frac{\epsilon}{r^{2}}\dot{t}\dot{r}+\frac{\epsilon^{2}}{r^{3}}
(1-kr)\dot{t}\dot{r}=0,
\end{eqnarray}
\begin{eqnarray}\nonumber
&&\ddot{r}-r(\dot{\theta}+\sin^{2}\theta\dot{\phi}^{2})+
\epsilon[\frac{1}{2r^{2}}(\dot{t}^{2}-\dot{r}^{2})+\dot{\theta}^{2}+\sin^{2}\theta
\dot{\phi}^{2}]\\\label{61}
&&-\frac{\epsilon^{2}}{2r^{3}}[(1+rk)(\dot{t}^{2}+\dot{r}^{2})-2kr^{3}
(\dot{\theta}+\sin^{2}\theta \dot{\phi}^{2})=0,
\end{eqnarray}
while the last two equations remain the same under these
perturbations. Here we use the same procedure as in the previous
section and obtain the same symmetries as for the electrically
charged black hole but in this case the re-scaling factor is
different. The energy re-scaling factor turns out to be
$(1-rk)=1-r\frac{Q^{2}}{M^{3}}$.

\section{Summary and Discussion}

In this paper, we have discussed the approximate symmetries of the
stringy charged black hole solutions. These solutions have the
isometry algebra $so(3){\oplus}R$ while the system of geodesic
equations have $so(3){\oplus}R{\oplus}d_{2}$. Firstly, we have
found the second-order approximate symmetries of the orbital
equation of motion. For the stringy magnetically charged solution,
there does not exist any non-trivial approximate symmetry
generator. Only exact and the first-order symmetry generators are
recovered as the second-order trivial symmetry generators. For the
electrically charged solution, the exact and first-order symmetry
generators are found to be the same as for the Schwarzschild
spacetime. However, for the second-order approximation, there
exist some non-trivial symmetry generators given by
Eqs.(\ref{35})-(\ref{41}).

Secondly, we have calculated symmetries of the geodesic equations
for both the spacetimes. For these charged solutions, the exact
and first-order approximate part of the symmetry generator turn
out to be the same as that for the Schwarzschild spacetime. For
the second-order approximation to symmetry generators, we get no
non-trivial symmetries. We have only recovered the lost
conservation laws as the second-order approximate conservation
laws. But unlike the Schwarzschild spacetime, from the perturbed
geodesic equations, we have obtained some re-scaling factor. The
re-scaling factor for the electrically charged solution is
$(1+4rk)=1+\frac{r\sinh^2{\alpha}}{m}$.

Using M$\o$ller's prescription, the energy distribution is given
by \cite{41}
\begin{equation}\nonumber
\frac{mr^{2}}{(r+2m\sinh^{2}\alpha)^{2}}[1+2(1-\frac{m}{r})\sinh^{2}\alpha].
\end{equation}
The re-scaling of force for the electrically charged solution
calculated through the pseudo-Newtonian formalism \cite{42} is
$$-\frac{\frac{m}{r^2}(1+\sinh^2\alpha-\frac{2m}{r}\sinh^2\alpha)}
{(1-\frac{2m}{r})(1+\frac{2m}{r}\sinh^2\alpha)}$$ which reduces to
the Schwarzschild solution for charge to be zero. Also, the
re-scaling factor of energy vanishes for $\alpha$ (charge) to be
zero (i.e. the Schwarzschild solution). It is mentioned here that
all the three expressions are $r$ dependent.

For the magnetically charged solution, the re-scaling factor turns
out to be $(1-rk)=1-r\frac{Q^{2}}{M^{3}}$. The re-scaling of force
by using the pseudo-Newtonian becomes
$$-\frac{\frac{M}{r^2}(1-\frac{Q^2}{2M^2})}{2r(1-\frac{2M}{r})(1-\frac{Q^2}{Mr})}.$$
The expression for energy calculated through Einstein's
prescription is \cite{43}
\begin{equation}\nonumber
E(r)=M-\frac{1}{2}r-\frac{M}{2Q^{2}}r^{2}-\frac{M^{2}}{2Q^{4}}r^{3}-
\frac{M^{3}}{2Q^{6}}r^{4}+O(r^{5}).
\end{equation}
In the limit of $Q=0$, it must reduce to energy for the
Schwarzschild spacetime but in the above expression, this limit
leads to $E(r)\longrightarrow \infty$. However, in the limiting
case, the re-scaling factor leads to the result of the
Schwarzschild spacetime. Further, energy calculated through
M$\o$ller's prescription is given by \cite{41}
\begin{equation}\nonumber
E(r)=\frac{(2M^{2}-Q^{2})r}{2(Mr-Q^{2})}
\end{equation}
which reduces to energy of the Schwarzschild spacetime for $Q=0$.
We would like to mention here that both the re-scaling factors
(obtained in our case) are $r$ dependent. However, it turns out to
be more simple evaluated by using the approximate symmetry as
compared to expressions evaluated through other approaches. Also,
for the electrically charged solution, it is seen that there exist
some non-trivial symmetries found through orbital equation of
motion. This fact is in contrast to the RN spacetime where no
non-trivial part of symmetry generator exists but the re-scaling
factor is independent of $r$.

In literature \cite{25,26}, a difference between the conservation
laws for full system of geodesic equations and single orbital
equation of motion is noted. We conclude that this difference also
holds for the stringy charged solutions. It is found that when
some symmetries are lost at one order (exact or the first-order
approximation) then they are recovered at the next order (at least
to the second-order) as trivial approximate symmetries.

It would be worthwhile to investigate the approximate symmetry
generators and the concept of energy re-scaling for the rotating
stringy black hole solutions which contain electric, magnetic and
both charges \cite{44,45}. For the complete understanding of this
procedure, it would be interesting to examine the colliding
gravitational waves \cite{46}. Work on these lines is in progress.

\vspace{0.3cm}

\renewcommand{\theequation}{A\arabic{equation}}
\setcounter{equation}{0}
\section*{Appendix A}

The symmetry generators of the Minkowski spacetime are given as
follows:
\begin{eqnarray}\label{14}
\textbf{X}_{0}&=&\frac{\partial}{\partial t},\quad
\textbf{X}_{1}=\cos\phi \frac{\partial}{\partial \theta}-\cot\theta
\sin\phi\frac{\partial}{\partial \phi},\\\label{15}
\textbf{X}_{2}&=&\sin\phi\frac{\partial}{\partial \theta}+\cot\theta
\cos\phi\frac{\partial}{\partial \phi},\\\label{16}
\textbf{X}_{3}&=&\frac{\partial}{\partial \phi},\\\label{17}
\textbf{X}_{4}&=&\sin\theta\cos\phi \frac{\partial}{\partial r}+
\frac{\cos\theta \cos\phi}{r} \frac{\partial}{\partial
\theta}-\frac{\csc\theta\sin\phi}{r}\frac{\partial}{\partial\phi},\\\label{18}
\textbf{X}_{5}&=&\sin\theta\sin\phi \frac{\partial}{\partial
r}+\frac{\cos\theta \sin\phi}{r}\frac{\partial}{\partial \theta
}+\frac{\csc\theta \cos\phi}{r}\frac{\partial}{\partial\phi},
\\\label{19}
\textbf{X}_{6}&=&\cos\theta\frac{\partial}{\partial r}-\frac{\sin
\theta}{r}\frac{\partial}{\partial\theta}, \\\label{20}
\textbf{X}_{7}&=&r \sin\theta \cos\phi\frac{\partial}{\partial
t}+t(\sin\theta \cos\phi\frac{\partial}{\partial r}+\frac{\cos\theta
\cos\phi}{r}\frac{\partial}{\partial\theta}\nonumber\\
&-&\frac{\csc\theta\sin\phi }{r}
\frac{\partial}{\partial\phi}),\\\nonumber
\textbf{X}_{8}&=&r\sin\theta \sin\phi\frac{\partial}{\partial
t}+t(\sin\theta \sin\phi\frac{\partial}{\partial r}+\frac{\cos\theta
\sin\phi}{r}\frac{\partial}{\partial\theta}\\\label{21}
&+&\frac{\csc\vartheta \cos\phi }{r}
\frac{\partial}{\partial\phi}),\\\label{22} \textbf{X}_{9}&=&r
\cos\theta\frac{\partial}{\partial
t}+t(\cos\theta\frac{\partial}{\partial r}-\frac{\sin
\theta}{r}\frac{\partial}{\partial\theta}).
\end{eqnarray}

\end{document}